\documentclass[floatfix,preprint,citeautoscript,amsmath,amssymb,aip,preprint]{revtex4}
\usepackage{dcolumn}
\usepackage{graphicx,setspace}
\usepackage{color}

\usepackage{graphicx}
\usepackage{dcolumn}
\usepackage{bm}

\begin{document}

\title{ Electronic structure and charge transfer excited states of a multichromophoric antenna }

\author{Luis Basurto$^1$, Rajendra R. Zope$^2$, and Tunna Baruah$^{1,2}$\footnote{Corresponding author: tbaruah@utep.edu}       }

\affiliation{$^1$ Computational Science Program, The University of Texas at El Paso, El Paso, TX 79958, USA }

\affiliation{$^2$Department of Physics, The University of Texas at El Paso, El Paso, TX 79958, USA }


\date{\today }

\begin{abstract}    

     The electronic structure of a multichromophoric molecular complex containing two of each borondipyrromethane dye, Zn-tetraphenyl-porphyrin,
bisphenyl anthracene and a fullerene are studied using density functional theory. The snowflake shaped molecule behaves like an antenna 
capturing photon at different frequencies and transferring the photon energy to the porphyrin where electron transfer occurs from the 
porphyrin to the fullerene.   
  Molecular structure of this large complex is first optimized using plane wave projector augmented wave methodology. Subsequent electronic 
structure calculations are performed using the real space methodology using an all electron pseudopotential basis set containing total 
of 12478 basis functions.
The results show that the HOMO and a state below the HOMO are primarily localized on one of the porphyrins while the LUMO resides 
mainly on the fullerene component of the complex.
The energies of the HOMO and LUMO states in the complex, as adjudged by the ionization potential and the electron affinity values,
show significant differences with respect to their values in participating subunits in isolation. We have systematically
examined the effect of structural strain and the presence of ligands on the ionization energy and the electron affinity.
Finally, we have calculated a few lowest charge transfer energies involving electronic transitions from a 
the porphyrin component to the fullerene subunit of the complex using the perturbative delta-SCF method.
Our predicted value of the lowest charge transfer excited state (1.67 eV) is comparable to the experimental 
estimate of the charge transfer energy of a similar complex. 

\end{abstract}

\maketitle

\section{Introduction}
 The organic heterojunction photovoltaics are often are designed as a donor-acceptor complex that consists of a p-type and an n-type semiconductor. 
These materials can be in molecular or polymeric form \cite{RFW:2221,RFW:2227,RFW:1459,RFW:1677,RFW:2236,RFW:1678}. However, often the absorption band 
is limited to that of the donor material. One way to overcome 
this limitation is to modify the electronic structure of the chromophore through chemical groups to broaden the 
absorption spectrum. Another way
that has been pursued by a few groups is to use an antenna-like construct \cite{antenna_Gust_1999,Panda20122601,doi:10.1021/ja055903c,doi:10.1021/ja063081t,doi:10.1021/jp073121v,CHEM:CHEM201002333,Gust_2009}.
   Such molecular 
antennas are made to mimic the action of biological antennas seen in plants. The function of biological antenna is to capture
solar energy at different wavelengths and funnel the energy to the reaction center. One such interesting artificial molecular antenna was   synthesized 
recently by Gust and co-workers \cite{Gust_2009}. This molecular antenna contains a  wheel shaped hexaphenylbenzene core where each of the phenyl 
rings is connected to a chromophore forming a hexad. The supramolecule contains
two of each of the chromophores : porphyrin (either H$_2$ or Zn), bis(phenylethynyl)anthracene (BPEA), and 
borondipyrromethane (BODIPY).   Both the BPEA and BODIPY units function by absorbing photons at different wavelength and subsequently
funneling the absorbed photon energy to one of the porphyrins. One advantage of such a construct is that together the BPEA and BODIPY widen the 
absorption band to the region where porphyrin absorption is weak. 
The BPEA moities absorb in 430-475 nm region which is between the porphyrin Soret and Q-bands. On the other hand BODIPY absorb
in the 475-530 nm and 330-430nm region. Thus the absorption range is quite extended for this complex.  Another advantage is 
that singlet-singlet energy transfer takes place from both the BPEA and BODIPY to the porphyrin.
Similar to the reaction center in natural light-harvesting systems, an electron transfer takes place from the porphyrin to the
fullerene.
The dynamics in the base porphyrin hexad without the fullerene moieties is slightly different 
from that of the Zn-porphyrin hexad.
In the absence of an acceptor moieties, an electron transfer takes place from the Zn-porphyrin to  BODIPY. 
However, in the presence of a C$_{60}$ molecule,
rapid electron transfer takes  place from the porphyrin to the C$_{60}$. The experiments were performed in  1,2 difluorobenzene and
2-methyltetrahydrofuran
both of which are polar solvents with 1,2 difluorobenzene ($\epsilon=13.8$) being more polar than 2-methyltetrahydrofuran($\epsilon=7.36$). 

The dynamics of the electron and energy transfer in this system makes it very interesting. The present article reports a density functional theory \cite{DFT1,DFT2} (DFT)
based electronic structure study of the hexad in its ground state and the lowest charge transfer state. 
We have recently developed a perturbative $\Delta$-SCF methodology 
to describe the charge transfer 
states of donor-acceptor systems. The initial applications on smaller systems showed excellent agreement with 
experiment \cite{JCP_TCNE}. In this article, we present a 
study of the lowest charge transfer state in the complex using the perturbative delta-SCF method.
In the next section we describe the computational methods followed by results and discussions of our study.

 \section{COMPUTATIONAL METHOD }

  All the calculations reported here are done using the density functional theory using the PBE exchange-correlation functional within the 
generalized gradient approximation (GGA) \cite{RFW:183}.  
The large size of the
complex presents challenges for quantum mechanical calculations. 
 To reduce the computational costs, 
the methyl groups are replaced by hydrogens resulting in a complex with 421 atoms. The structure optimization of the heptad was
carried out using PAW pseudopotentials as implemented in the VASP code \cite{PAW1,PAW2,VASP1,VASP2,VASP3,VASP4}. The VASP optimized structure is then used 
to derive all the results reported here using the  NRLMOL code \cite{NRLMOL1,NRLMOL2,NRLMOL3,NRLMOL4}. The NRLMOL code has previously been 
used to study large light-harvesting  systems \cite{RFW:1089,ISI:000246173100030,ISI:000236397600003}. In this work, we have used both the 
all-electron formalism and pesudopotential formalism wherever all-electron approach was difficult to apply due to the large size of the molecule.
we used a mixed all-electron and pseudopotential approach for the ground state calculations.
 In the ground state calculation, all-electron basis is used for the hydrogens and the zinc atoms whereas for all the other types of atoms pseudopotential
given by Bachelet, Hammam and Schluter is used \cite{BHS}. 

The numbers of the primitive Gaussians, s-type, p-type and d-type functions along with the 
range of the exponents for each type of atoms are given in table \ref{table_basis}.  
All the contracted basis functions for a given atom are derived from the same set of primitive Gaussians. 
The mixed  basis set contained 
12478 basis functions in total. The all-electron basis set used in this work were 
optimized for PBE-GGA functional \cite{RFW:181}. In a similar way, the pseudopotential basis is 
also optimized for the BHS pseudopotential. 
The basis set used here is larger than the typical 6-311G basis set used for moderate size molecules.
Our efforts to  perform  calculations at the all-electron level
with the 6-311G\*\* basis showed numerical instability forcing the convergence criteria to be  reduced.
Because of the numerical errors, we performed the ground state
calculations with the mixed pseudopotential and all-electron approach using the NRLMOL basis functions only. 
To reduce the computational cost, only spin-restricted calculations are done.
\begin{table}[h]
\begin{tabular}{lccccl}
\hline
Atom & s-type & p-type & d-type & Primitives & Exponent range \\
\hline
Pseudo-potential basis \\
B    &   4    & 4      & 3      & 5                   & 4.16       - 0.069 \\
C    &   4    & 4      & 3      & 6                   & 5.65       - 0.077 \\
N    &   4    & 4      & 3      & 6                   & 8.55       - 0.104 \\
O    &   4    & 4      & 3      & 7                   & 1.05 x10   - 0.100 \\
F    &   4    & 4      & 3      & 7                   & 1.45 x10   - 0.131 \\
All-electron basis \\
C    &   5    & 4      & 3      & 12                   & 2.22 x 10$^4$       - 0.077 \\
N    &   5    & 4      & 3      & 13                   & 5.18 x 10$^4$       - 0.094 \\
H    &   4    & 3      & 1      & 6                   & 7.78 x 10  - 0.075 \\
Zn   &   7    & 5      & 4      & 20                  & 5.00 x $10^6$ - 0.055 \\
\hline
\end{tabular}
\caption{ The numbers of s-, p-, d-type contracted functions, number of primitive Gaussian and the range of the
Gaussian exponents used for each atom.}
\label{table_basis}
\end{table}
   The intramolecular excitation energies for the electronic transitions occurring on the same part of the molecule
are studied using the time dependent theory as implemented in Gaussian09\cite{g09}.  
The charge transfer excitations are not well described by the TDDFT unless specially optimized 
range corrected functionals are used. However, perturbative delta SCF method as illustrated in our recent article\cite{JCP_TCNE,Baruah_JCTC}
provides satisfactory description of charge transfer excitations. The method reproduces experimental values 
of charge transfer excitation energies 
for a database of tetracyanoethylene-hydrocarbon complexes and other model organic photovolatic complexes within 
0.3 eV or less. This set also includes a porphyrin fullerene complex which are the units involving charge transfer
in the present study.
The notable feature
of this method is that it maintains the orthogonality constraint between the ground state and excited state
Slater determinantal wavefunctions.  This method uses a perturbative approach to determine the excited state
orbitals and density and does not contain any empirical or system dependent parameters.  
The method has been previously used to study the charge transfer excitations in a few fullerene porphyrin dyads\cite{JCP_Dyad,JCP_endon}
and carotene-porphyrin-fullerene triad\cite{Baruah_JCTC}. Its predictions on C70 porphyrin\cite{JCP_Dyad} are consistent with 
recent many-body Green's function GW study\cite{Blase}. For the details of the method and its performance 
we refer reader to  our recent articles Ref. \onlinecite{Baruah_JCTC,JCP_TCNE}. Since the excited state calculations 
require larger memory, we have used a triad cutout 
of the heptad for the excited state calculations. The calculations on both the ground and excited states of 
the triad cutout was carried out at the all electron level. We have further verified that the charge transfer 
excited state energy
using the pseudopotential differs from the all-electron approach by only 0.04 eV.

\section{RESULTS AND DISCUSSION}
  
\begin{figure}
\includegraphics[width=8.5cm]{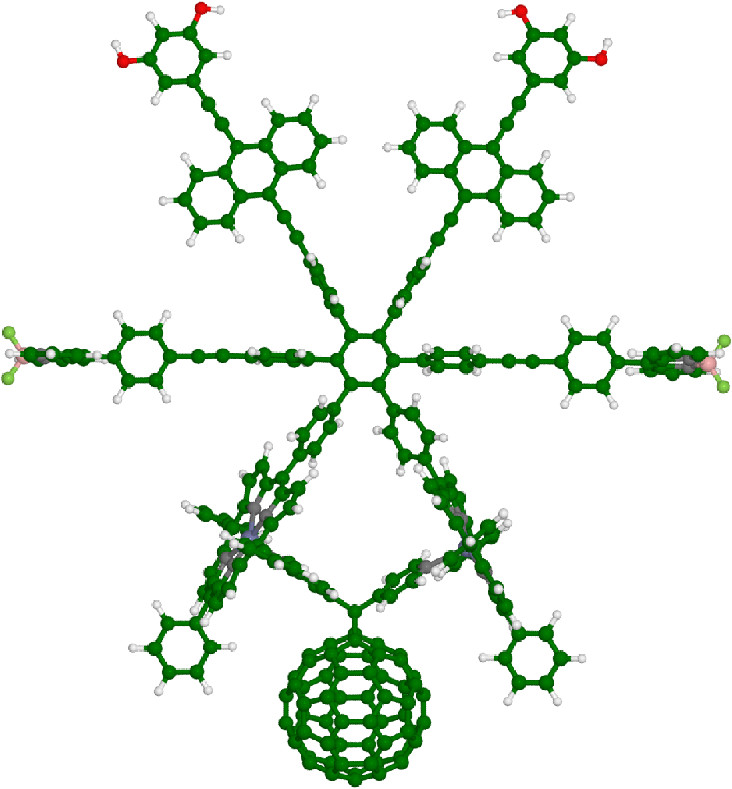}
\caption{ (Color online) The structure of the heptad complex.}
\label{str}
\end{figure}

  The heptad molecule contains a hexaphenylbenzene core where the phenyl rings lie at nearly 90$^o$ angle to the central benzene ring.
The planes of the anthracenes are at $\sim$ 90$^o$ angle to the 
phenyl ring of the core whereas those of the BODIPY are in plane with their corresponding phenyl rings. On the other hand the porphyrins 
are strained such that 
the rings connected to the porphyrins are distorted from 90$^o$ angle. 
 The DFT optimized structure of the heptad molecule is 
shown in Fig. \ref{str}. 
The electron donor (zinc-tetraphenyl porphyrins) and acceptor(fullerene) moieties are connected through a pyridine with a separation of 
roughly 6.8 \AA between the Zn ion at the porphyrin 
center and the nearest fullerene surface. 
The two pyridines are connected through a single 
carbon on top of a 6:6 bond of the fullerene.  
 The fullerene molecule is rigidly wedged between the two porphyrins such that torsions 
of the fullerene-porphyrin linker is
unlikely to occur. Since
the structure of the electron donor-acceptor part of the complex is rigid, the possibility of isomerization  in the 
presence of a solvent is much lower \cite{Gust_2009}.

\begin{figure}
\includegraphics[width=10.6cm]{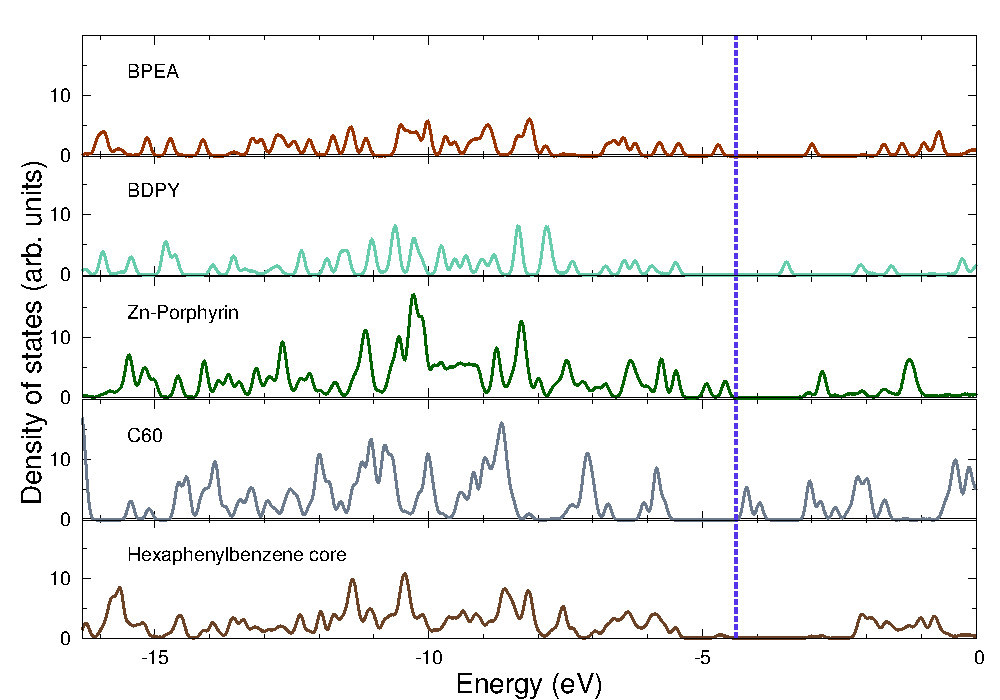}
\vspace{0.4cm}
\caption{ (Color online) The density of states of the heptad molecule. The site projected density of states on the various components are shown.}

\label{DOS}
\end{figure}

The calculated density of states of the gas-phase heptad is shown in Fig. \ref{DOS}. The fermi energy is marked as a straight line 
in the plots.
The site-decomposed  DOS is also presented for the various distinct units of the supramolecule - the hexaphenyl benzene core, 
Zn-porphyrin, BODIPY, BPEA and the fullerene. Since the DOS belonging to the two components of Zn-porphyrin, BODIPY, and BPEA are identical, 
only one set of DOS for each of these units are shown. The pyridine units are considered as part of the porphyrin moieties for plotting the DOS. 
Since the molecule has  two symmetrically placed porphyrins, the 
highest two occupied molecular orbitals, HOMO and HOMO-1 of the heptad, are localized on the porphyrins are nearly degenerate. 
They are predominantly located on one of the porphyrins with some orbital density also on the other counterpart.
The fullerene LUMOs form the lowest three LUMO of the heptad which split into a doublet and a singlet with an energy  spacing   
of 0.24 eV between them at DFT GGA level.
Thus the lowest charge transfer occurs from one of the porphyrins to the fullerene. 
The HOMO of BPEA lies in between the two occupied Gouterman orbitals \cite{Gouterman1961} of the porphyrins whereas the HOMO of the BODIPY is 
lies deeper than the porphyrin
Gouterman orbitals. 
The HOMO of the BODIPY is about 0.3 eV lower than the HOMO-2 of the Zn-porphyrin.
However, the LUMO of the BODIPY is only slightly higher than fullerene LUMOs 
as seen from the Fig. \ref{DOS}. 
The HOMO-LUMO separation for these different units of the heptad as given by the KS-DFT 
ground state calculation are  
about 1.5 eV for Zn-porphyrin, 1.5 eV for 
fullerene, 2.0 eV for BODIPY and 1.3 eV for BPEA. These gaps are underestimated in the KS-DFT 
calculations due to the self-interaction errors 
and missing derivative discontinuities in DFT functionals but it qualitatively shows the relative ordering of the orbitals belonging to different units.  
The ordering seen in the orbitals localized on different units  supports the fact in the absence of the fullerene moieties, charge transfer
occurs from the porphyrin to the BODIPY which was noted by Gust et al. \cite{Gust_2009}.

  The dipole moment of the heptad in its ground state is small due to its symmetrical structure. Our calculated value for the dipole 
is 3.98 Debye which points from the fullerene to the hexaphenyl-benzene core.

\begin{figure}
\includegraphics[width=8.5cm]{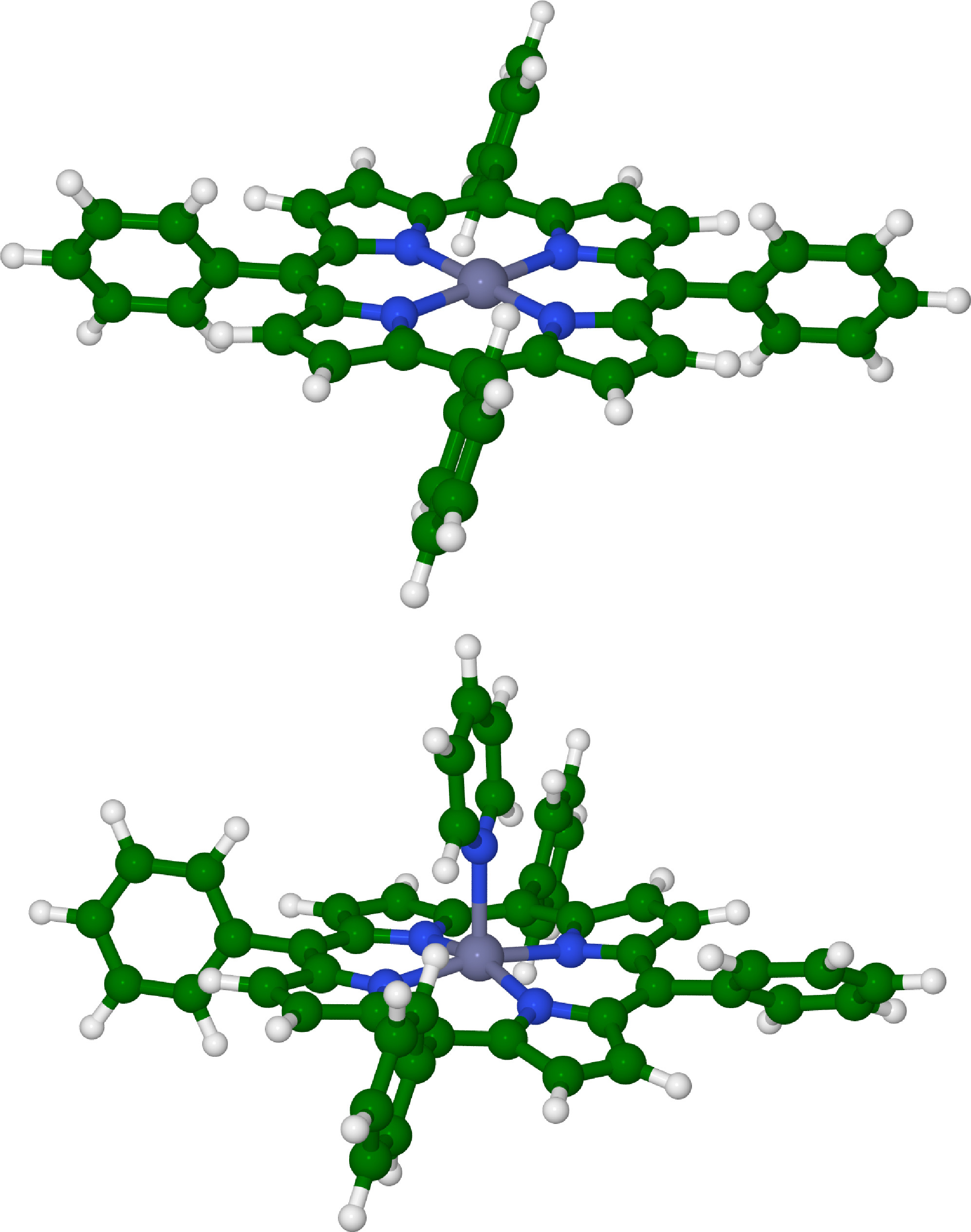}
\caption{ (Color online) The change in the Zn porphyrin structure due to the axial pyridine ligand.}
\label{porphyrin}
\end{figure}

 In the heptad molecule the pyridine ligands attached to the fullerene binds to the zinc ions in the ZnTPPs.  
Thus the Zn ions are  in five-fold coordination
in this compound. The porphyrins in the heptad molecule are strained.
We have relaxed the structure of a 
 free ZnTPP molecule with 
an axial pyridine ligand. The ligand  changes the structure
such that the phenyl rings are twisted (Cf. Fig. \ref{porphyrin}).
The porphyrin plane itself also puckers out.  Such puckering of the porphyrin plane was 
reported earlier in a number of studies \cite{ZnTPP-pyridine,Walker_ZnTPP,Bobrik_ZnTPP}.
Relaxation of the free porphyrin structure as present in the heptad reduces the energy of the pyridine 
ligated Zn tetrapenylporphyrin by 0.51 eV.
These results not only bring out the impact of the pyridine ring on the electronic structure of the porphyrin 
but are also indicative of the strain on the pyridine-porphyrin units in the heptad molecule. 
One of the effects of the structural distortion is to reduce
the HOMO-LUMO gap.
 We have employed both the TDDFT method and also a perturbative delta-self-consistent field method at the all-electron 
spin-polarized formalism to calculate the
excitation energies of the pyridine-ligated ZnTPP and free ZnTPP without the axial ligand. 
The first singlet excitation of a free ZnTPP with no axial ligand is calculated to be at 2.03 eV using the 
perturbative delta-SCF. 
This  energy is close to the value of 2.01 eV for the lowest singlet 
excited state obtained from TDDFT.
For an isolated relaxed ZnTPP with the axial pyridine, the calculated energy of the lowest singlet excited state  
using the perurbative delta-SCF method is 1.84 eV.  
The other singlet excitations in free pyridine-ligated-porphyrin 
are 1.90 eV (H$\longrightarrow$ L+1), 2.17 (H-1$\longrightarrow$L), 2.21 eV (H-1 $\longrightarrow$L+1), and 2.99 eV (H$\longrightarrow$L+2).
Since mixed characters of excited states are not 
well reproduced by this method, we have also calculated these excited state energies using TDDFT method. The TDDFT calculation was 
carried out using the Gaussian09 code\cite{g09}. The TDDFT calculation shows that the lowest singlet is a mixing of H$\longrightarrow$L, H$\longrightarrow$L+2,H-1$\longrightarrow$L+1 
excitations with an excitation energy of 1.99 eV. The next singlet is at 2.01 eV.
The mixing of the states in the pyridined-porphyrin is different from the free ZnTPP. In free ZnTPP the lowest
singlet is comprised of H$\longrightarrow$L, H$\longrightarrow$L+1, H-1$\longrightarrow$L, and H-1$\longrightarrow$L+1 excitations. 
 The lowest  singlet excited state of pyridine-ligated-Zn-porphyrin in the heptad molecule 
was estimated at 2.03 eV by Gust et al. \cite{Gust_2009}.  The ZnTPP Q band is reported to show a red-shift of $\sim$ 15 nm
due to the appended pyridine coordinated to the metal atom \cite{ZnTPP-pyridine-absorp,ZnTPP-pyridine-absorp2,ZnTPP-pyridine-C60}

  Our calculated value for the porphyrin local excitation in the (ZnTPP)$_2$-C$_{60}$ is 1.75 eV and 
the excitation from the porphyrin HOMO-1 to porphyrin LUMO occurs at 2.13 eV. These energies are slightly lower than
those for the free pyridined-porphyrin mostly due to strain. This was confirmed by calculating the lowest excitation
in the strained pyridine ligated porphyrin. Mixing of the states similar to that 
predicted by TDDFT method for free pyridined-porphyrin is likely 
increase the energy of the lowest singlet excited state. The TDDFT calculations on triad could not be carried out due to its large
size.

\begin{figure}
\includegraphics[width=8.5cm]{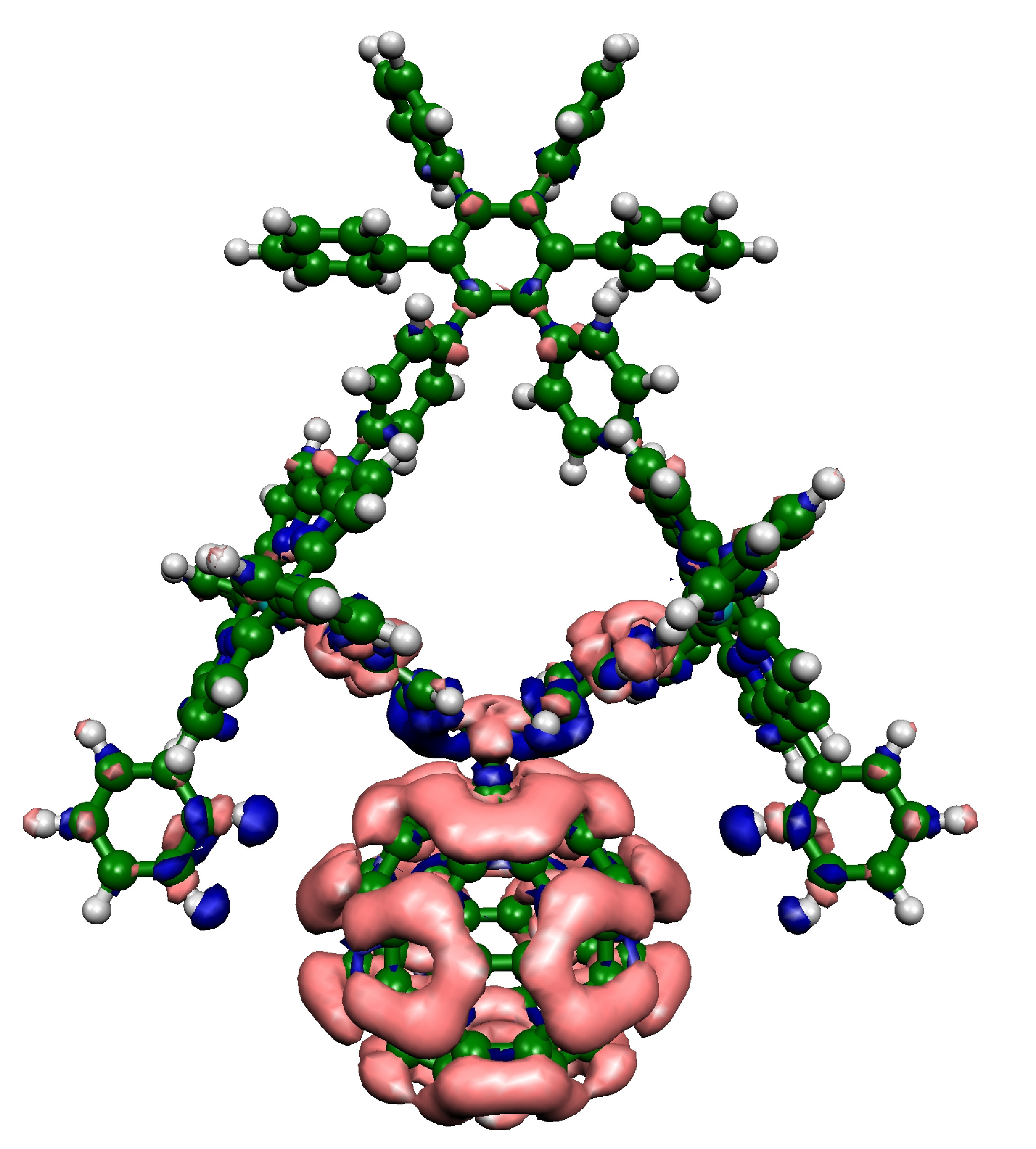}
\caption{ (Color online) The density difference between the anionic and neutral state.}
\label{anion}
\end{figure}

  The calculated ionization potential (IP) of the triad cutout is also much smaller compared to the free ZnTPP. This 
happens possibly due to the fact that the
HOMO is delocalized  with substantial density on the second porphyrin and also due to the
presence of the axial ligation to the pyridine connecting the porphyrin to the C$_{60}$. 
The HOMOs of the two porphyrins are degenerate. Our calculated ionization potential of a single 
porphyrin with an axial
pyridine on top shows that the IP changes from 6.34 eV for a free ZnTPP to 6.00 eV for the pyridine-ligated-porphyrin. 
Lack of experimental data on ultraviolet photoelectron spectra of  pyridine-ligated porphyrin hinders a direct comparison.
The electrochemical measurements are done in solution in which the choice of the solvent is important.
Experimentally, a change of $\sim$ 0.11 eV was reported in the oxidation potential of Zn-tetraphenyl porphyrin in pyridine
in electrochemical measurements in solution \cite{Gust_1993,kadis_1981}.  
Strain on the porphyrin plane, similar to that present in the heptad, further reduces it to 5.86 eV.
The calculated value of the  IP of the (ZnTPP)$_2$-C$_{60}$ at all-electron level is 5.54 eV. 
The IP of the full heptad molecule using mixed pseudopotential and all-electron approach differs only slightly at 5.49eV.
The HOMO level is spread over both the porphyrins although it is mostly localized on one.
This spread may raise the HOMO energy further up thereby reducing the ionization energy.
The electron affinity of the (ZnTPP)$_2$-C$_{60}$ is also higher (2.94 eV) than that for an 
isolated C$_{60}$ molecule (2.68 eV). 
These changes are much larger than the change seen in non-bonded 
C$_{60}$-ZnTPP complexes in cofacial arrangement or endon orientations \cite{JCP_Dyad, JCP_endon}.
Furthermore, experimental measurement of reduction potential of a dipyridine C$_{60}$ model reports a change of 10-40 meV only \cite{Gust_2009}. 
Similar change was noted for single  pyridine connected to a fulleropyrolidine \cite{ZnTPP-pyridine-C60}.
Our calculations on a free dipyridine-C$_{60}$ molecule but with same structure as in the heptad shows that the verical electron affinity 
increases from free C$_{60}$ by about 0.09 eV. Relaxation of the structure of this dipyridine-C$_{60}$ resulted in decrease of electron affinity
to 2.62 eV. A plot of the difference of the density in the neutral and the anionic state shows density distribution both on the 
C$_{60}$ as well as on the pyridine rings (Cf. Fig. \ref{anion}). Possible polarization effects on the porphyrin components may lower 
the energy of the anionic state. 

 We have used the perturbative $\Delta$-SCF method to  determine the lowest charge transfer excitation energy 
of the heptad which 
 occurs from the porphyrins to the fullerene. In the perturbative $\Delta$-SCF method, the occupied orbitals are expaned in the space of
 unoccupied 
orbitals using a pertubative approach. 
As mentioned earlier, it applications to small donor-acceptor organic conjugates\cite{JCP_TCNE} as well as to supramolecular 
carotenoid-porphyrin-C$_{60}$ triad \cite{Baruah_JCTC} and porphyrin-fullerene dyads\cite{JCP_Dyad}
show that the method can be reliably used to determine the charge transfer excitation energies for systems with vanishing 
transition dipole moments.
The method is computationally as expensive for a given excited state as the Kohn-Sham DFT for the ground state. However, the excited state calculations 
require larger memory since the information about the ground state Hamiltonian is retained.
Because of the large memory requirement for the calculation of the excitation energies, we have used a smaller
triad model of the heptad containing only the donor and acceptor moities as shown in Fig. 2. The geometry of the triad was 
not optimized to mimic the geometry of that part of the heptad. This part of the heptad will be referred as triad cutout hereafter.
The HOMO of the triad cutout is on the porphyrins and the LUMO is on the fullerene. 
Since this model system is smaller in size, we used an all-electron approach to calculate the excited states.
Our calculated CT excitation energy
from the HOMO on the Zn-porphyrin to the LUMO of the fullerene in the triad is 1.67 eV. 
We find that the excited state energies differ only by 0.04 eV if we use a pseudopotential basis instead of an all-electron basis.
 Experimental estimate of the CT energy on the full heptad molecule is  made from the reduction potential of a model C$_{60}$-dipyridine molecule and 
oxidation potential of pyridined-ZnTPP. This energy is estimated to be
1.37 eV by Gust et al.
\cite{Gust_2009}. On another similar bis-porphyrin-fullerene triad , the experimental value of the CT excitation from porphyrin to C$_{60}$ 
is found to be 1.46 eV \cite{Gust_2011}. 
 The linkers connecting the porphyrins to the fullerene in the bis-porphyrin-fullerene triad in Ref. \onlinecite{Gust_2011} 
are quite different. The effect due to the axial pyridine ligands is hence missing and therefore a 
direct comparison between our calculated value and experimental estimate is not possible. 
We have also calculated the charge transfer excitation from a porphyrin HOMO-1 to fullerene LUMO. Energy of this CT excited state 
is 2.07 eV.  
  The plots of orbital densities show a low lying virtual bridge state situated on the pyridine linkers. Our calculations
show that excitation from  HOMO to the pyridine bridge state (Fig. \ref{homo_lumo5}) leads to another CT state at 2.82 eV which shows that this state is unlikely
to participate in any charge transfer transition resulting from porphyrin Q-band. 
The calculated excitation energies do not account for any structural reorganization of the complexes.
The ionic relaxation in the triad or heptad is likely to be small due to the highly constrained structures of 
the complexes. 
 The electrochemical measurements 
in experiments are carried out in polar solvents such as methyltetrahydrofuran where the solvent reaction field 
can further stabilize the excited state. The dipole moment of the ground state of the triad cutout 
is 2.48D and in the CT excited state it increases to 36D. The distance between the Zn ion to nearest fullerene 
surface distance is about 6.8 \AA in this complex which explains the
dipole moment of the CT state.

\begin{figure}
\includegraphics[width=8.5cm]{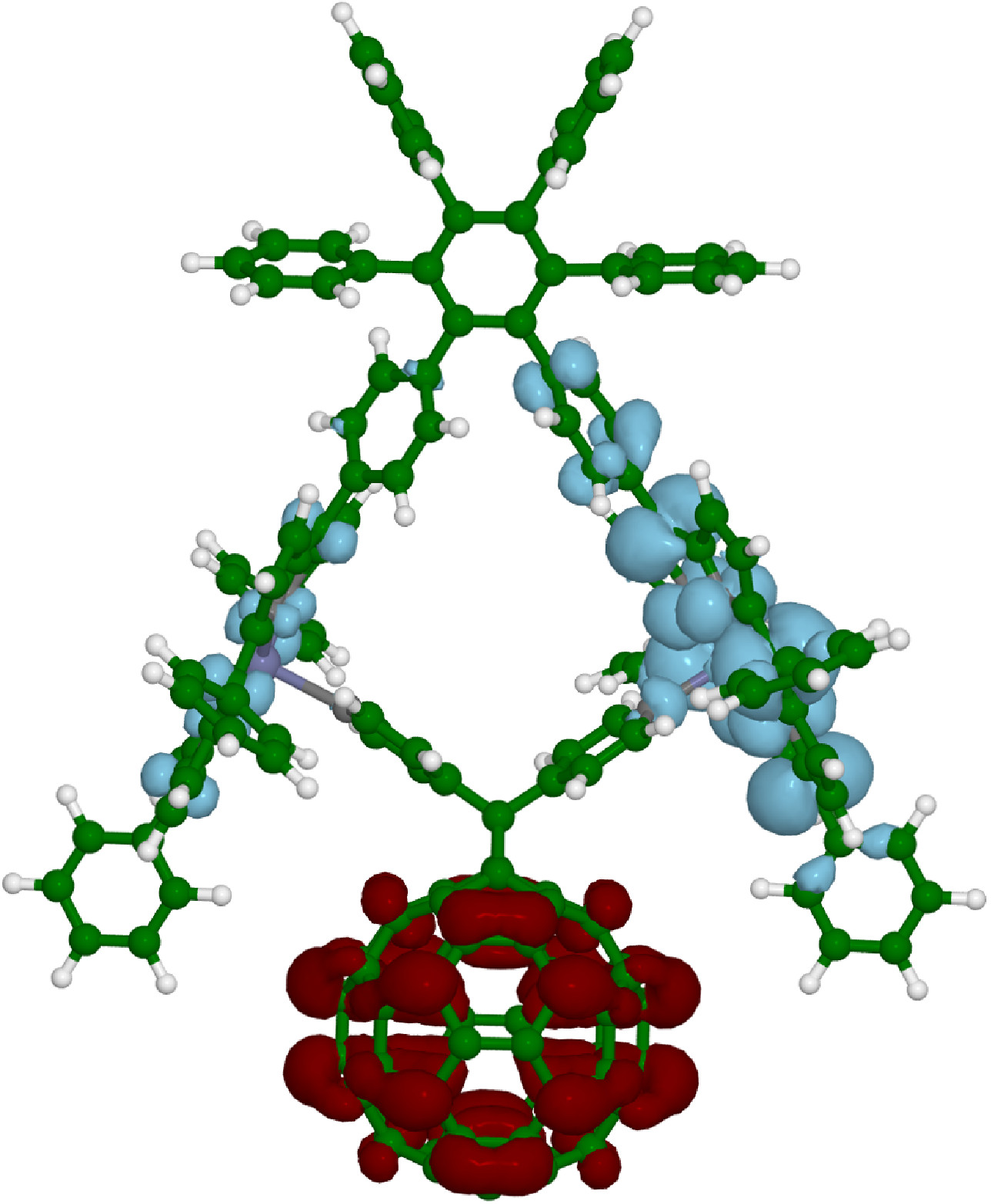}
\caption{ (Color online) The lowest HOMO to LUMO charge transfer state. }
\label{homo_lumo}
\end{figure}

\begin{figure}
\includegraphics[width=8.5cm]{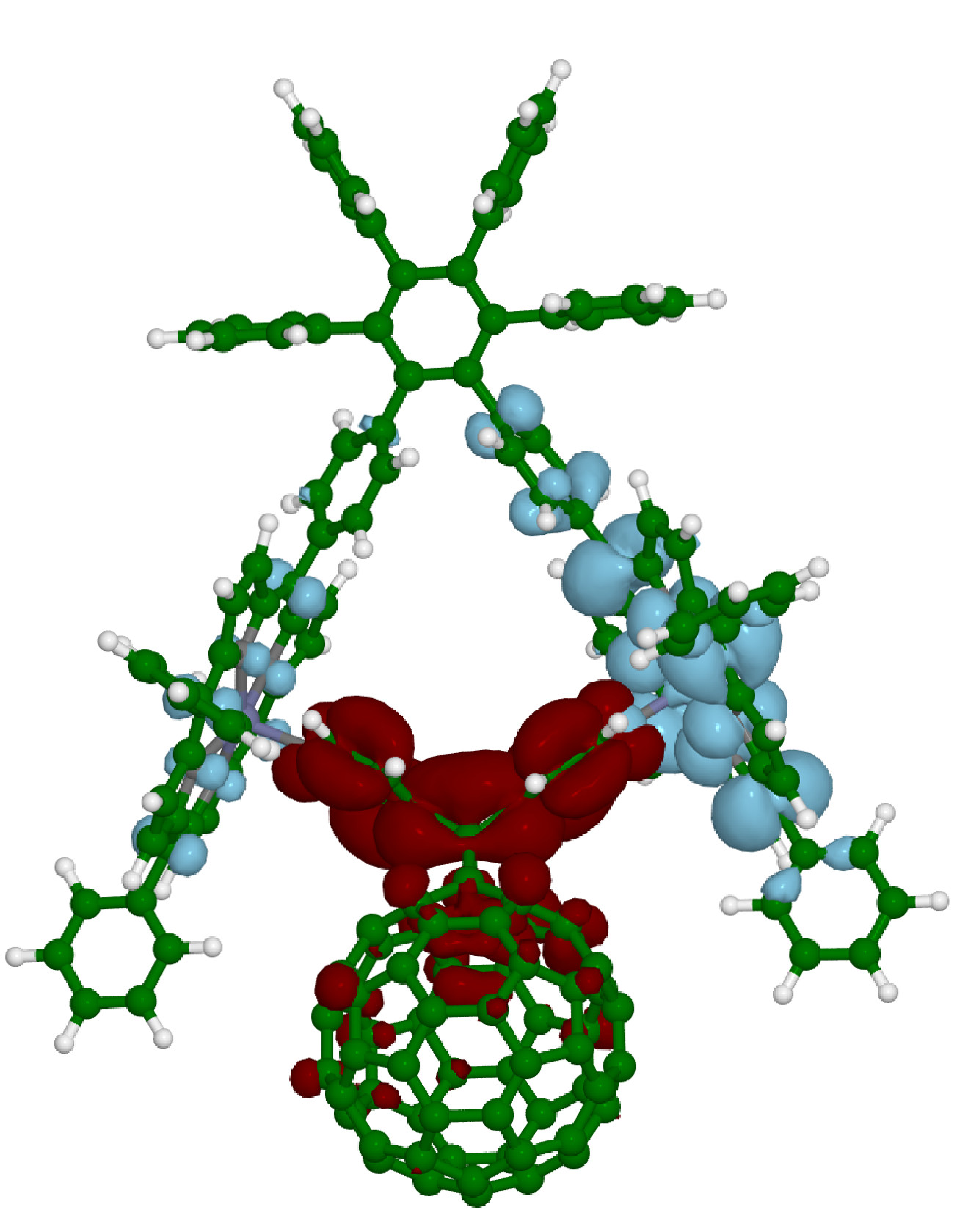}
\caption{ (Color online) A ZnTPP to bridge charge transfer state. }
\label{homo_lumo5}
\end{figure}

In summary, we have studied the electronic structure using DFT of a  multichromophoric molecular heptad that behaves like an antenna. In its ground state, the 
highest occupied molecular orbital is located mostly on one of the porphyrins and the lowest unoccupied MO is on the fullerene. We find that the BODIPY HOMO 
lies deep compared to other components of the heptad. 
In agreement with experimental observation of Gust et al. \cite{Gust_2009} our calculations indicate that in absence of the fullerene,
the electron electron transfer would occur from the Zn-porphyrin to the BODIPY. The ionization potential and electron affinity of the heptad is quite different than that for a
ZnTPP or ZnTPP with pyridine ligand. The strain on the porphyrins, presence of the axial pyridine ligands, and the delocaliztion
of the HOMO orbital contributes to the reduction of the ionization potential. The electron affinity of the haptad also significantly 
differs for the heptad from an isolated C$_{60}$ or isolated dipyridine-C$_{60}$ molecule possibly due to polarization effects.
Our calculated value of the HOMO-LUMO charge transfer energy for a representative bis-porphyrin-fullerene triad cutout is 1.67 eV.
The dipole moments of such CT states are high 36D.
In this molecule the reorganization of the porphyrins and fullerenes is likely to be small due to the
structural constraints.
The difference between the calculated value and the experimental energy is likely due to the
effects of polar solvents which will stabilize the excited state.

  This work was funded by the Division of Chemical Sciences, Geosciences, and 
Biosciences, Office of Basic Energy Sciences of the U.S. Department of Energy
through Grant DE-SC0002168. It is a pleasure to acknowledge helpful discussions 
with Mark Pederson and Marco Olguin. The support for computational time at Texas Advanced Computing Center by the NSF through grant 
TG-DMR090071  and at National Energy Research Scientific Computing center is gratefully acknowledged.


\end{document}